\documentclass[10pt]{iopart}
\usepackage{graphicx}
\usepackage{cite}
\usepackage{varioref}

\begin{document}

\title[Semiclassical ionization dynamics of the hydrogen molecular ion]
 {Semiclassical ionization dynamics of the hydrogen molecular ion in an
  electric field of arbitrary orientation}
\author{T Bartsch and T Uzer}
\address{Center for Nonlinear Science, School of Physics, Georgia Institute
  of Technology, Atlanta, GA 30332-0430, USA}
\ead{bartsch@cns.physics.gatech.edu}

\begin{abstract}
Quasi-static models of barrier suppression have played a major role in our
understanding of the ionization of atoms and molecules in strong laser
fields. Despite their success, in the case of diatomic molecules these
studies have so far been restricted to fields aligned with the molecular
axis. In this paper we investigate the locations and heights of the
potential barriers in the hydrogen molecular ion in an electric field of
arbitrary orientation. We find that the barriers undergo bifurcations as
the external field strength and direction are varied. This phenomenon
represents an unexpected level of intricacy even on this most elementary
level of the dynamics. We describe the dynamics of tunnelling ionization
through the barriers semiclassically and use our results to shed new light
on the success of a recent theory of molecular tunnelling ionization as
well as earlier theories that restrict the electric field to be aligned
with the molecular axis.
\end{abstract}

\submitto{\jpb}

\section{Introduction}
\label{sec:Intro}

The description of the electronic dynamics in the hydrogen molecular ion
was one of the important issues in the early days of quantum mechanics and
was fundamental to the development of quantum chemistry
\cite{Pauli22,burrau}.  Recent progress in different fields of physics
has thrust this seemingly museumish, if venerable problem back into the
focus of current research: the study of the nonlinear, nonperturbative
interaction of matter with intense, ultrashort pulses
\cite{Bandrauk94,Posthumus01,Bandrauk03} and of the fascinating internal
dynamics of Rydberg plasmas \cite{pillet1,dutta,killian1,killian2}. Both
demand the investigation of the electron motion under the influence of two
stationary Coulomb centres (TCC) and an external static electric field: On
the one hand, quasistatic models of laser-matter interaction that neglect
the time-dependence of the laser field have here been surprisingly
successful \cite{corkum1,krause,corkum2,Bandrauk03}.  On the other hand, in
a cold plasma the background ions are almost stationary, so that their
Coulomb fields are static to a good approximation \cite{feldbaum}.

In the present paper, we initiate an investigation of the classical and
semiclassical mechanics of the electron motion under the influence of two
Coulomb centres and an external field of arbitrary orientation. We focus on
a description of the saddle points of the relevant potential because they
regulate the ionization of the molecule \cite{Uzer02}. The locations and
heights of the saddles depend on the orientation of the external field and
on its strength. Apart from this expected smooth change, we find
bifurcation points where additional potential saddles are created or
collide and are destroyed. The occurrence of bifurcations even on this most
elementary level of the dynamics indicates that the full dynamics must be
much more intricate than appears at first sight. The heights of the
potential saddles determine the possibility or impossibility of classical
over-the-barrier ionization. They depend strongly on the direction of the
field. After a detailed discussion of the onset of over-the-barrier
ionization, we turn the dynamics of tunnelling ionization, which has
recently been the subject of intense study
\cite{Zuo95,Bandrauk00,Tong02}. Using methods of modern nonlinear dynamics,
we propose a dynamical mechanism that underlies the success of recent
theory of molecular tunnelling ionization \cite{Tong02} that is fashioned
after the well-known \emph{Ammosov-Delone-Kra\u\i nov\/} (ADK) theory
\cite{Ammosov86}.  The techniques we apply are based on classical mechanics
that obeys scaling laws (see section~\ref{sec:Potential}). For this reason,
although we focus our exposition on the laser ionization of molecules, the
results are equally valid for the much larger internuclear distances
relevant to the physics of Rydberg plasmas.  At the same time, our methods
generalize straightforwardly to more complicated systems, such as the
electronic motion under the influence of more than two Coulomb centres,
possibly in the presence of electric and magnetic fields.

In contrast to atoms, which are spherically symmetric in the absence of
external fields, a diatomic molecule possesses a preferred molecular
axis. The reaction of the molecule to an electric field will depend not
only on the field strength, but also on its direction. In addition, an atom
in an external field retains rotational symmetry around the field axis,
which renders the electronic dynamics effectively two-dimensional. In a
molecule any external field that is not aligned with the molecular axis
breaks all continuous symmetries, leads to a strong coupling of all three
degrees of freedom and thus induces a significantly more complicated
dynamics. Because it has been shown both experimentally
\cite{Normand92,Dietrich93} and theoretically \cite{McCann92,Aubanel93}
that a diatomic molecule in a strong laser field experiences a torque that
tends to align its axis with that of the electric field, and also because
it avoids the intricacies due to the inclusion of an additional degree of
freedom, investigations of the quasistatic dynamics have mainly been
restricted to the two-dimensional case of parallel axes
\cite{Codling89,Posthumus95,Posthumus96a,Plummer96,Smirnov97,Smirnov98}. In
many cases, the ionization of the molecule was even described in terms of a
one-dimensional model \cite{Codling89,Posthumus95,Posthumus96a}. However,
first investigations of the misaligned case \cite{Plummer97} show that its
dynamics is remarkably different from the case of aligned axes.

Classical and semiclassical methods such as the \emph{field
ionization--Coulomb explosion} (FICE) model
\cite{Codling89,Posthumus95,Posthumus96a} have served well to illuminate
the dynamics of molecular ionization in a static field. Their virtue is
that they can provide an intuitive picture of the essential dynamics, even
in situations where many coupled degrees of freedom are involved and that
are intractable by exact quantum mechanical methods. For this reason it is
highly desirable to make their power available to the description of the
general case which, due to the complete breaking of symmetries, can be
expected to be much more complex. In this paper, we will embark on a
description of the electronic dynamics in the field of two stationary
Coulomb centres (TCC) together with a static electric field of arbitrary
strength and arbitrary orientation. Given the success of the FICE model
\cite{Codling89,Posthumus95,Posthumus96a} for the collinear configuration
of non-hydrogenic and even heteronuclear molecules, we expect our results
to be valuable beyond the hydrogen molecular ion for a wide range of more
complicated molecules.

We will focus on the description of the saddle points of the TCC system in
an electric field, i.e.~the unstable equilibrium points where the electron
can be at rest. They are the critical bottlenecks where ionization takes
place. It has been shown recently \cite{Uzer02} that, using the methods of
modern nonlinear dynamics, one can compute ionization rates of atoms in
external fields from a detailed description of the saddle
points. Therefore, after laying out some general properties of the TCC
potential in an external field in section~\ref{sec:Potential}, we will, in
section~\ref{sec:Saddles}, start our investigation with the identification
of the saddle points the potential possesses. Already in this seemingly
elementary first step, we will encounter the richness the dynamics of our
system exhibits: We will find that the potential saddles undergo
bifurcations as the field strength and field direction are varied, so that
for any given external field either two or four saddle points can
coexist. We will also see that the saddles differ in their indicess, i.e.~in
the number of unstable directions attached to them.

The most fundamental property of a saddle point is its height. The height
determines if classical over-the-barrier ionization can take place (namely,
when the saddle is lower than the electron energy) or is forbidden. It
therefore marks the onset of ionization. We will discuss the heights of the
different saddles as function of field strength, field direction and
internuclear distance in section~\ref{sec:Thresh}.

If the energy of a bound electron is lower than the classical ionization
threshold, the molecule can still ionize by means of tunnelling
\cite{keldysh,delone,corkum1}. This process is commonly described by the
ADK model \cite{Ammosov86}, which uses the ionization rate of a hydrogen
atom or hydrogen-like ion in a static external field. In that system,
ionization takes place predominantly along the axis of the external
electric field, so that the description of the ionization process reduces
to a one-dimensional tunnelling calculation. The ADK model has
recently been generalized by Tong \etal \cite{Tong02} to explain the
tunnelling ionization of diatomic molecules. These authors explain the
ionization suppression found experimentally for a wide range of molecules
by taking into account the anisotropy of the bound-state molecular wave
functions. They neglect, however, the coupling of the different degrees of
freedom that is absent in atomic ionization, but relevant in a molecule.
In section~\ref{sec:Tunnel} we will analyze the tunnelling dynamics for the
hydrogen molecular ion in a misaligned electric field from the point of
view of nonlinear dynamics and thereby elucidate the reason why the
approach of Tong \etal could nevertheless succeed. Our approach will also
clarify the success of theories that assume the electric field to be
aligned with the molecular axis
\cite{Codling89,Zuo95,Posthumus95,Posthumus96a,Plummer96,Smirnov97}.

\section{The TCC potential in an electric field}
\label{sec:Potential}

In the following, we will discuss the dynamics of an electron under the
combined influence of two stationary nuclei of charge $Z$ and a homogeneous
static electric field. We will choose the coordinate system so that the
external field is in the $xz$-plane and the nuclei are lying on the
$z$-axis, at a distance $c$ on both sides of the origin. With these
conventions, the electronic dynamics can be described by the Hamiltonian,
in atomic units,
\begin{equation}
  \label{Ham}
  H=\frac{{\bi p}^2}{2} + V({\bi x})
\end{equation}
with the potential
\begin{equation}
\label{Potential}
  \fl
  V({\bi x}) = -\frac{Z}{\left(x^2+y^2+(z-c)^2\right)^{1/2}}
               -\frac{Z}{\left(x^2+y^2+(z+c)^2\right)^{1/2}}
	       -Fz\cos\phi - Fx\sin\phi \;.
\end{equation}
Here, $F$ is the strength of the external field.  The angle $\phi$ between
the field and the negative $z$-axis uniquely specifies the field
direction. Notice that we have taken $V$ to denote the potential energy of
an electron with negative unit charge. It is the negative of the
electrostatic potential.

Apart from the external field strength and direction, the
potential~(\ref{Potential}) and the Hamiltonian~(\ref{Ham}) depend on the
nuclear charge $Z$ and the internuclear distance $2c$. Those dependences
can be removed by introducing the scaled quantities
\begin{eqnarray}
\label{scaling}
    \tilde{\bi x}={\bi x}/c \;, \qquad
    \tilde{\bi p}=(c/Z)^{1/2}{\bi p} \;, \nonumber\\
    \tilde F=c^2 F/Z \;, \nonumber\\
    \tilde H = cH/Z\;, \qquad \tilde V=cV/Z \;.
\end{eqnarray}
In terms of these scaled variables, the scaled Hamiltonian reads
\begin{equation}
  \label{ScalHam}
  \tilde H=\frac{\tilde{\bi p}^2}{2} + \tilde V(\tilde{\bi x})
\end{equation}
with the scaled potential
\begin{equation}
\label{ScalPotential}
  \fl
  \tilde V(\tilde{\bi x}) = 
      -\frac{1}{\left(\tilde x^2+\tilde y^2+(\tilde z-1)^2\right)^{1/2}}
      -\frac{1}{\left(\tilde x^2+\tilde y^2+(\tilde z+1)^2\right)^{1/2}}
      -\tilde F\tilde z\cos\phi - \tilde F\tilde x\sin\phi \;.
\end{equation}
The nuclear charges have been scaled to 1 and the internuclear distance to
2. It is this scaled form of the dynamics that we will use in most of what
follows, returning to the unscaled variables as necessary.

Notice that the Hamiltonian~(\ref{ScalHam}) is symmetric with respect to a
reflection in the $xz$-plane, so that this plane is invariant under the
dynamics. Furthermore, for $\phi=90^{\circ}$ the external field is oriented
along the $x$-axis and there is an additional symmetry with respect to
reflection in the $xy$-plane. In the general case, a reflection in that
plane takes an external field oriented at an angle $\phi$ from the negative
$z$-axis into a field oriented at an angle $180^{\circ}-\phi$, so that the
dynamics in these two situations agree.

\section{The potential saddles of the TCC system in an electric field}
\label{sec:Saddles}

In this section we will identify and describe the equilibrium points of the
potential~(\ref{ScalPotential}) where $\nabla\tilde V=0$.
Section~\ref{ssec:EqLoc} describes their locations. In a second step, in
section~\ref{ssec:EqClass}, we will compute the Hessian matrix of the
potential in these points to classify them as potential maxima, minima or
saddles.

\subsection{The location of equilibria}
\label{ssec:EqLoc}

The task of finding the stationary points of the combined potential can be
rephrased as finding those points where the Coulomb field caused by the two
centres is equal in magnitude and opposite in direction to the applied
external field. We will therefore, in this section, restrict ourselves to a
discussion of the pure TCC potential and locate those points where the TCC
electric field has a given strength and direction. For symmetry reasons it
is obvious that these points can only lie in the $xz$-plane spanned by the
molecular axis and the direction of the external field. We can therefore
further restrict our discussion to the TCC field strength in that
plane. Due to the sign conventions taken in~(\ref{Potential}), the external
field makes an angle $\phi$ with the negative $z$-axis and, if
$\phi$ is chosen in the interval $0<\phi<180^{\circ}$, has a negative $x$
component. As a consequence, the TCC field balancing the external field
must have a positive $x$ component and make and angle $\phi$ with the
positive $z$-axis. This is the range of parameters to which we will
restrict our attention in the following.

\begin{figure}
\centerline{\includegraphics[width=.48\textwidth]{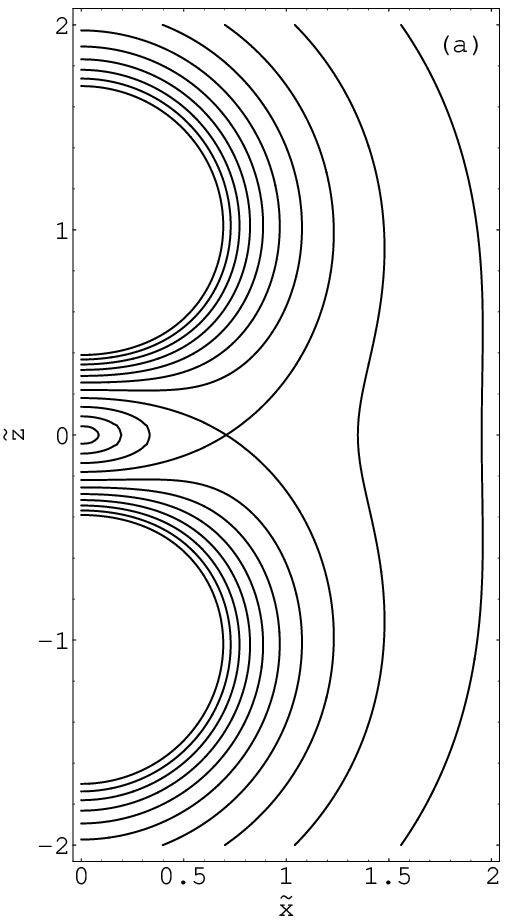}\hfill
            \includegraphics[width=.48\textwidth]{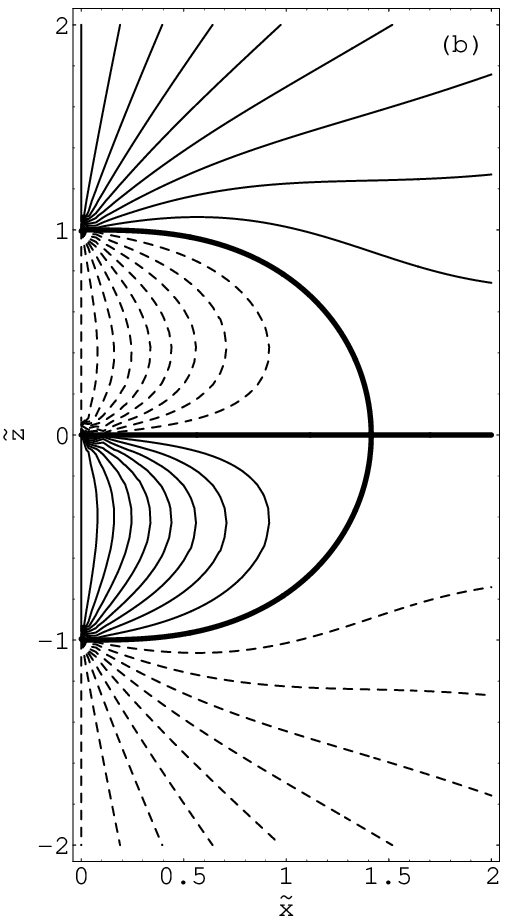}}
\caption{\label{fig:TccField}
  (a) Contour lines of the absolute value of the TCC electric field
  strength. (b) Contour lines of the angle $\phi$ between the
  TCC electric field and the positive $\tilde z$-axis. Thick lines:
  90-degree angle line, thin solid lines: angle lines $\phi<90^{\circ}$,
  dashed lines: $\phi>90^{\circ}$.}
\end{figure}

Let us first discuss the absolute value of the TCC field strength,
disregarding the field direction. The variation of field strength in the
$xz$-plane is illustrated in figure~\ref{fig:TccField}(a). Regions of high
field strength are located in the vicinity of either nucleus. There, the
field caused by the close-by nucleus is much stronger than that of the
remote nucleus, so that the lines of constant field strength are nearly
concentric circles around the nucleus.  At the other extreme, the field
strength is zero at the midpoint between the nuclei, i.e.~at $x=z=0$, as
well as infinitely far from the centres.  For sufficiently low field
strengths, therefore, there is a field strength line surrounding the
midpoint and a second strength line surrounding both centres.

Thus, although the topology of field strength lines is different for high
and for low field strength, for any given field strength there are two
disjoint contours. The transition between the different topologies occurs
for the critical field strength $\tilde F_{\rm c}$ where the disjoint
contours merge and intersect at right angles. It can be inferred from
figure~\ref{fig:TccField}(a) that $\tilde F_{\rm c}$ is the maximum field
strength taken on the $x$-axis. From this observation we can calculate
that $\tilde F_{\rm c}=4/(3\sqrt{3})=0.7698$.

We now turn to a description of the direction of the two-centre Coulomb
field, which is illustrated in figure~\ref{fig:TccField}(b). We first look for
those locations where the field is perpendicular to the molecular axis. Due
to symmetry, it is clear that this is the case on the $\tilde x$-axis. In
addition, 90-degree angle
lines emanate from either nucleus and join the $\tilde x$-axis
at $\tilde x=\sqrt{2}$. As can be seen from figure~\ref{fig:TccField}(b), they
partition each quadrant of the $\tilde x\tilde
z$-plane into an ``inner'' region close to $\tilde x=\tilde z=0$ and an
``outer'' region at large $\tilde x$ and $\tilde z$. The angle lines for
angles $\phi<90^{\circ}$ fill the outer region of the quadrant $\tilde z>0$
and the inner region of the quadrant $\tilde z<0$. Symmetrically, the angle
lines for $\phi>90^{\circ}$ fill the inner region of the quadrant $\tilde
z>0$ and the outer region of the quadrant $\tilde z<0$. Thus, any angle
line for $\phi\neq 90^{\circ}$ has two disjoint parts. The 90-degree angle
lines themselves correspond to the critical value where the two parts of
the angle lines connect and change their topology.

Comparing figures~\ref{fig:TccField}(b) and~\ref{fig:TccField}(a), it
becomes clear that as we follow the outer part of any angle line from the
nucleus out to infinity, the field strength will decrease monotonically
from infinity to zero. As a consequence, any field strength is assumed
exactly once. If we apply an arbitrary external field, we will find a
unique equilibrium point on the corresponding outer angle line where the
Coulomb field exactly cancels the external field. We call it the
outer saddle point.

On the inner angle lines, the situation is more complicated. For
sufficiently small field angles $\phi$, the field strength along the inner
angle line will increase monotonically from zero to infinity as the line is
followed from midpoint to the nucleus. Therefore, in this
case we will find a single equilibrium on the inner angle line that will
turn out to be an inner saddle point.

\begin{figure}
\centerline{\includegraphics[width=.6\textwidth]{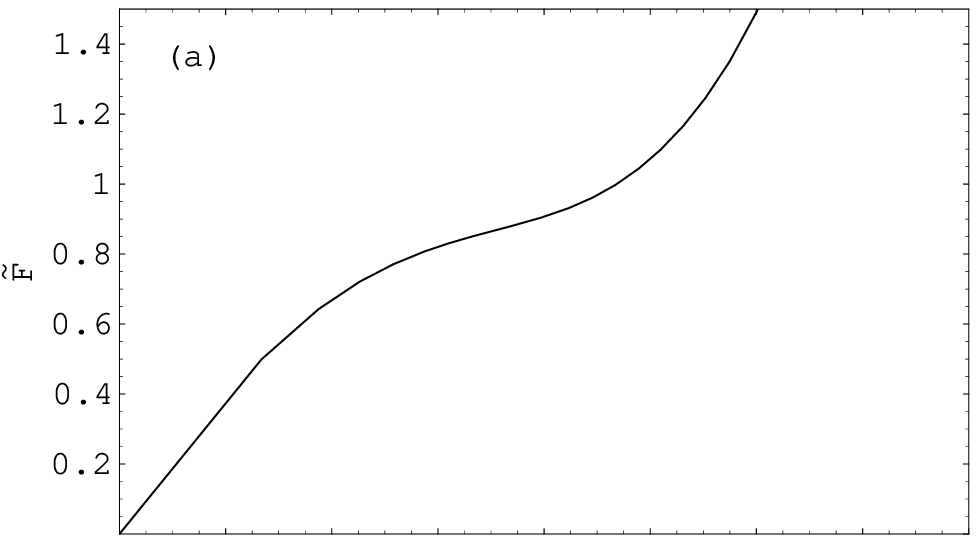}}
\centerline{\includegraphics[width=.6\textwidth]{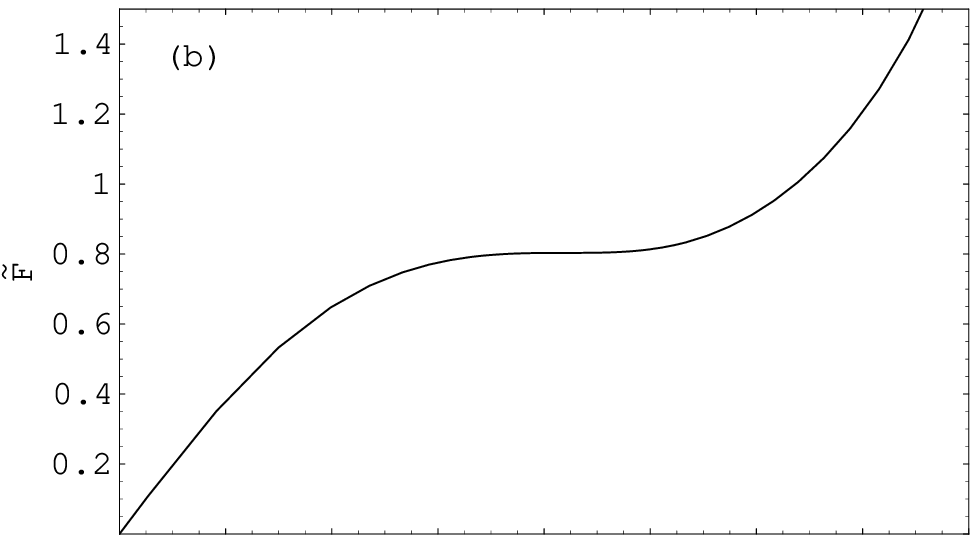}}
\centerline{\includegraphics[width=.6\textwidth]{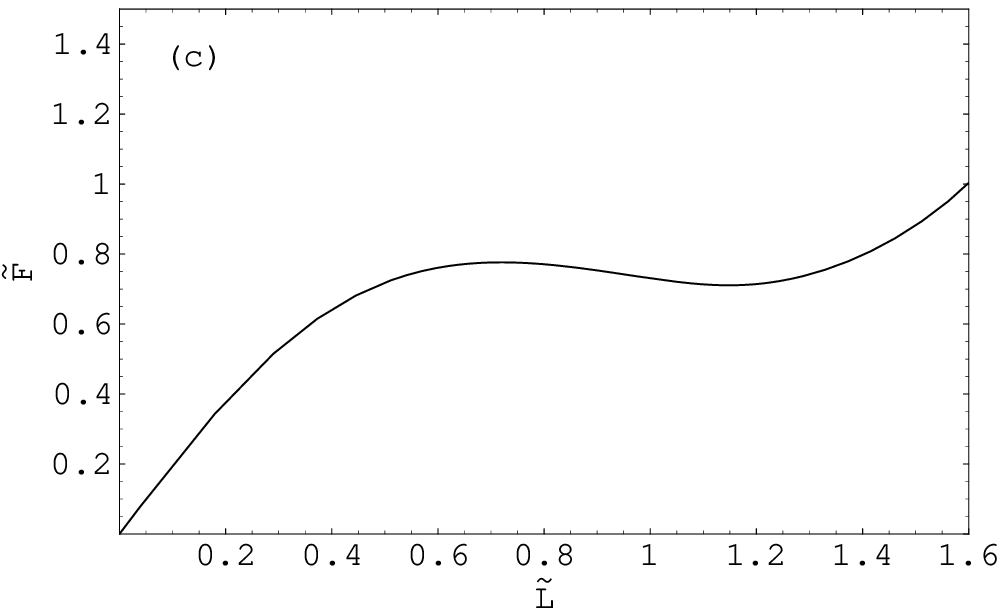}}
\caption{\label{fig:FInner}
  TCC field strengths measured along the inner
  angle lines for field angles (a) $\phi=75^{\circ}$, (b) $\phi=\phi_{\rm
  min}=81.6014^{\circ}$, (c) $\phi=86^{\circ}$. The parameter $\tilde L$ is
  the arc length of the angle line, measured from the midpoint between the
  nuclei.}
\end{figure}

\begin{figure}
\centerline{\includegraphics[width=.6\textwidth]{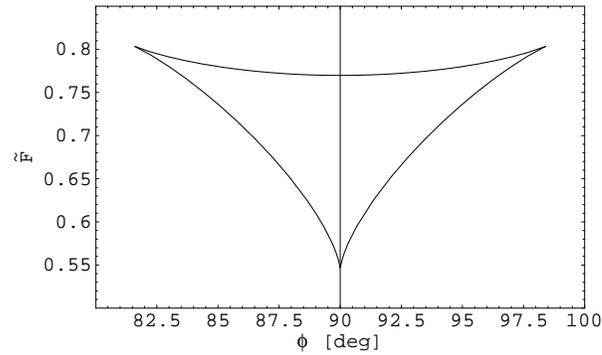}}
\caption{\label{fig:FMinMax}
  Minimum and maximum scaled field strengths $\tilde F_{\rm min}(\phi)$ and
  $\tilde F_{\rm max}(\phi)$ bounding the parameter region where three
  inner equilibria exist.}
\end{figure}

For field angles $\phi$ close to $90^{\circ}$, however, the field strength
along the inner angle line approximates the field strength along the
90-degree line. It rises from zero at the midpoint to a maximum $\tilde
F_{\rm max}(\phi)\approx\tilde F_{\rm c}$, then decreases to a minimum
$\tilde F_{\rm min}(\phi)$ and finally increases to infinity at the
nucleus. This behaviour is illustrated in figure~\ref{fig:FInner}. It is
found for angles in the interval $\phi_{\rm min}<\phi<\phi_{\rm max}$,
where $\phi_{\rm min}=81.6014^{\circ}$ and $\phi_{\rm
max}=180^{\circ}-\phi_{\rm min}=98.3986^{\circ}$. As a consequence, if the
field angle is chosen in this interval, there is a unique inner equilibrium
if the field strength is outside the interval $\tilde F_{\rm
min}(\phi)<\tilde F<\tilde F_{\rm max}(\phi)$, but there are three inner
equilibria if the field strength is within that interval.  The domain of
field strengths and field angles where multiple inner equilibria occur is
depicted in figure~\ref{fig:FMinMax}.

\begin{figure}
\centerline{\includegraphics[width=.6\textwidth]{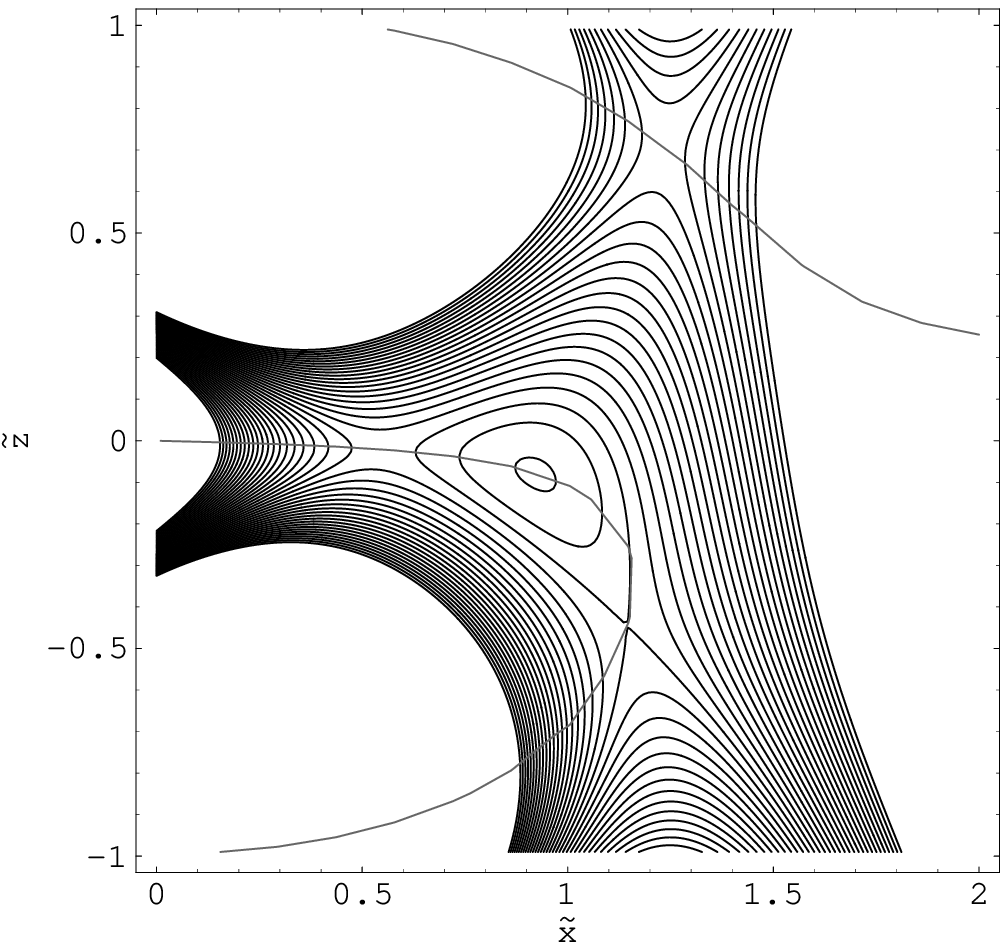}}
\caption{\label{fig:FourEq}
  Contour plot of the total potential for the TCC system in an external
  electric field for a configuration with three inner equilibria. The
  scaled field
  strength is $\tilde F=0.74$, the angle with the molecular axis
  $\phi=87^{\circ}$. The $87^{\circ}$-angle lines are indicated for
  clarity.}
\end{figure}

For a field configuration where three inner equilibria exist, the potential
in the $\tilde x\tilde z$-plane is shown in figure~\ref{fig:FourEq}. It
exhibits a saddle, a maximum, and a saddle along the inner angle line.
As the external field is varied and leaves the region of multiple
equilibria in figure~\ref{fig:FMinMax}, bifurcations of the equilibria must
occur. Specifically, as the upper critical field strength $\tilde F_{\rm
max}$ is approached from below, the maximum will collide with the saddle to
its left in the figure and they will annihilate, forming what in the
language of Catastrophe Theory \cite{Poston78,Castrigiano93} is called a
``fold'' catastrophe. Similarly, as the lower critical field strength
$\tilde F_{\rm min}$ is approached, the maximum will collide and annihilate
with the saddle below in the figure.

The most degenerate field configuration arises for the extremal angles
$\phi_{\rm min}$ and $\phi_{\rm max}$ where the lower and upper critical
field strengths coincide. For these parameter values, all three inner
equilibria coalesce. If we follow the potential along the angle line in
this situation, we find a minimum of fourth order. In the parlance of
Catastrophe Theory, this scenario is called a ``cusp'' catastrophe. It also
occurs for $\phi=90^{\circ}$ if the field strength is chosen such that the
saddles on the two outer 90-degree lines approach the $\tilde x$-axis and
there coalesce with a maximum on that axis. This bifurcation forms the
lower corner of the ``triangle'' in figure~\ref{fig:FMinMax} that bounds
the region of three inner equilibria. In all three corners of that
triangle, its edges meet tangentially, as is predicted by the theory of the
cusp catastrophe \cite{Poston78,Castrigiano93}.

\subsection{The classification of equilibria}
\label{ssec:EqClass}

Once we have found the equilibria of the TCC system in an external electric
field, which we have achieved in section~\ref{ssec:EqLoc}, we must classify
them according to the number of unstable degrees of freedom the dynamics
possesses in their neighbourhood. That number is given by
the number of negative eigenvalues of the Hessian determinant. We call it
the index of the equilibrium point \cite{Castrigiano93}. Note that it
suffices to discuss the behaviour of the potential in the $\tilde x\tilde
z$-plane of the three-dimensional potential. The motion in the $\tilde y$
degree of freedom decouples from the dynamics in the symmetry plane in the
harmonic approximation and is always stable: As we move away from the
plane, the distances to both nuclei increase and the potential goes
up. Thus, an equilibrium that is found to be a potential saddle in the
plane is a saddle of index one of the three-dimensional
potential. Similarly, a maximum of the planar potential corresponds to a
saddle of index two and a minimum of the planar potential to a minimum
(with index zero) of the three-dimensional potential.

To classify the equilibria of the planar potential, note that the (scaled)
planar Hessian determinant
\begin{equation}
\label{Hessxz}
  {\rm Hess}(\tilde x,\tilde z)= \det \left(
     \begin{array}{cc}
       \frac{\partial^2 \tilde V}{\partial^2 \tilde x } &
       \frac{\partial^2 \tilde V}{\partial\tilde x\partial\tilde z} \\
       \frac{\partial^2 \tilde V}{\partial\tilde z\partial\tilde x} &
       \frac{\partial^2 \tilde V}{\partial^2 \tilde z }
     \end{array}
     \right)
\end{equation}
of the potential $\tilde V$ is negative at a saddle and positive at a
maximum or minimum. Furthermore, since the second derivatives of the linear
external potential are zero, we can evaluate the Hessian of the TCC
potential instead of the total potential.

\begin{figure}
\centerline{\includegraphics[width=.6\textwidth]{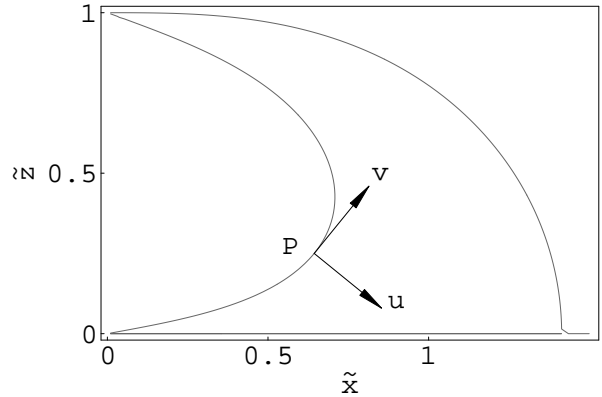}}
\caption{\label{fig:Coords}
  The coordinate system used for the calculation of the Hessian
  determinant. See text for a description.}
\end{figure}

To compute the Hessian at a point $P$, we introduce a Cartesian
coordinate system $(u,v)$ such that the $v$-axis is tangential to the angle
line through $P$ (see figure~\ref{fig:Coords}). Using the fact that the TCC
electric electric field strength $\tilde{\bi F}_{\rm TCC}=\nabla \tilde
V_{\rm TCC}$, we can rewrite the Hessian~(\ref{Hessxz}) as
\begin{equation}
\label{Hessuv}
  {\rm Hess}(P) = \det \left(
    \begin{array}{cc}
      \frac{\partial F_u}{\partial u} &
      \frac{\partial F_u}{\partial v} \\
      \frac{\partial F_v}{\partial u} &
      \frac{\partial F_v}{\partial v}
    \end{array}
    \right) \;,
\end{equation}
where
\begin{equation}
  \label{Fuv}
  F_u=\tilde F\sin\psi\;,\qquad F_v=\tilde F\cos\psi
\end{equation}
are the $u$ and $v$ components of the scaled TCC electric field strength and
$\psi$ is the angle between the electric field and the positive
$v$-axis. If $\alpha$ denotes the angle between the $v$- and $\tilde z$-axes,
$\psi$ is given by $\psi=\phi-\alpha$. Due to our choice of the $v$-axis
along the angle line, we have
\begin{equation*}
  \left.\frac{\partial \psi}{\partial v}\right|_P =0 \;,
\end{equation*}
so that the Hessian~(\ref{Hessuv}) with~(\ref{Fuv}) simplifies to
\begin{equation}
  \label{Hess}
  {\rm Hess}(P) = \tilde F\,\frac{\partial \tilde F}{\partial v}
                   \,\frac{\partial\psi}{\partial u}
                = \tilde F\,\frac{\partial \tilde F}{\partial v}
                   \,\frac{\partial\phi}{\partial u} \;.
\end{equation}

Referring back to figure~\ref{fig:TccField}(b), we see that in the inner region
shown $\partial\phi/\partial u$ is negative, because the $u$-axis points
from an angle line $\phi>90^{\circ}$ toward the 90-degree line. We thus
obtain the result that the Hessian of the potential at an equilibrium point
$P$ is negative, i.e.~$P$ is a saddle point, if and only if the derivative
of the field strength along the angle line is positive. That field strength
has already been discussed above. It is illustrated in
figure~\ref{fig:FInner}. (The orientation of the angle line chosen there is
the same as the orientation of the $v$-axis in figure~\ref{fig:Coords}.)
The field strength is monotonically increasing along the angle line
whenever there is a single inner equilibrium, which therefore is a
saddle. For angles and field strengths where three inner equilibria exist,
the first and the third encountered along the angle line are in a region of
increasing field strength and are therefore saddles, whereas the second is
either a maximum or a minimum of the potential. As long as the angle
between the positive $v$-axis and the Coulomb field is less then
$90^{\circ}$, which is certainly the case in the part of the angle line
that closely follows the $\tilde x$-axis, one can further show that the
potential has a maximum along the angle line if $\partial \tilde F/\partial
v<0$. Thus, the second equilibrium is found to be a maximum. By a similar
argument, it can be shown that the index of the outer saddle is always one.

\section{Classical ionization thresholds}
\label{sec:Thresh}

Having found the locations of the saddle points of the TCC potential in an
electric field, we now turn to a discussion of their heights, which
ultimately determine the possibility of classical over-the-barrier
ionization. We will focus, in particular, on the dependence of the barrier
heights upon the field angle. Before a discussion of general field
orientations, it will be useful to consider the extremal cases where the
electric field is oriented either parallel or perpendicular to the
molecular axis.

\subsection{Barriers in a perpendicular field}
\label{ssec:PPerp}

The perpendicular field configuration is special in that in addition to the
reflection symmetry in the $xz$-plane there is another symmetry with
respect to the $xy$-plane perpendicular to the molecular axis. As a
consequence, saddle points must either lie in that plane (i.e.~on the
$x$-axis) or occur in symmetrical pairs.

\begin{figure}
\centerline{\includegraphics[width=.6\textwidth]{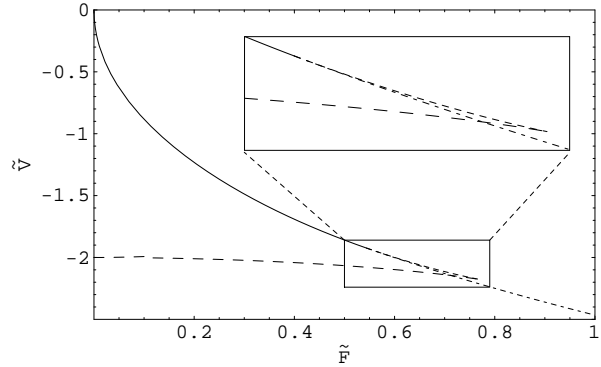}}
\caption{\label{fig:Height90}
  The heights of the potential saddles as a function of the electric field
  strength for $\phi=90^{\circ}$. Solid line: outer saddle, long-dashed
  line: inner saddle, short-dashed line: saddle of index two. All of these
  saddles are located on the $x$-axis. Dash-dotted line: symmetric pair of
  saddles away from the $x$-axis. The inset shows an enlarged view of the
  bifurcation region.}
\end{figure}

If the scaled field strength is low, there are an inner and an outer saddle
on the $x$-axis. With increasing field strength, the inner saddle moves
outward and the outer saddle inward. At $\tilde F=\tilde
F_1=(2/3)^{3/2}=0.5443$ the outer saddle crosses the point $\tilde
x=\sqrt{2}$ where the outer 90-degree angle lines reach the $x$-axis. At
this point, it turns into a saddle of index two, and a symmetric pair of
saddles of index one on the outer 90-degree lines is born. As the field
strength increases further, the latter move toward the nuclei along the
angle lines. The saddle of index two collides and annihilates with the inner
saddle at $\tilde F=\tilde F_2=4/(3\sqrt{3})=0.7698$. This leaves only the
pair of symmetric saddles at high field strength. Because the 90-degree
lines form the boundary between inner and outer angle lines, these saddles
can at will be classified as either ``inner'' or ``outer''
saddles. Figure~\ref{fig:Height90} illustrates this sequence of events. It
also shows that for most field strengths the inner saddle is lower that the
outer saddle. As in the case of an axial field
\cite{Codling89,Posthumus95,Posthumus96a,Smirnov97}, the inner saddle
allows the electron to switch from one nucleus to the other, not to
ionize. Thus, the onset of above-threshold ionization is, as in the axial
configuration, determined by the height of the outer saddle.

\subsection{Angle-dependence of the barrier heights}
\label{ssec:BarrAngle}

\begin{figure}
\centerline{\includegraphics[width=.6\textwidth]{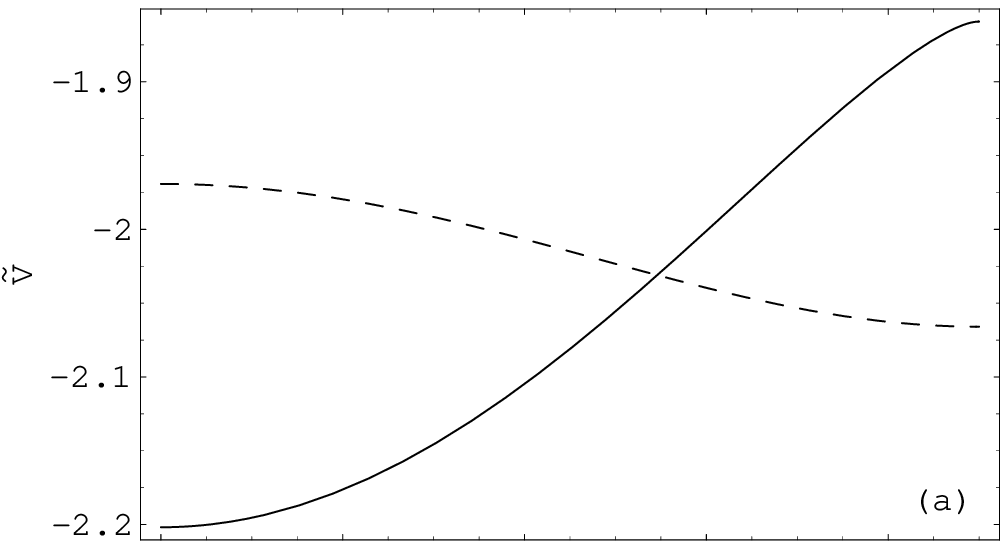}}
\centerline{\includegraphics[width=.6\textwidth]{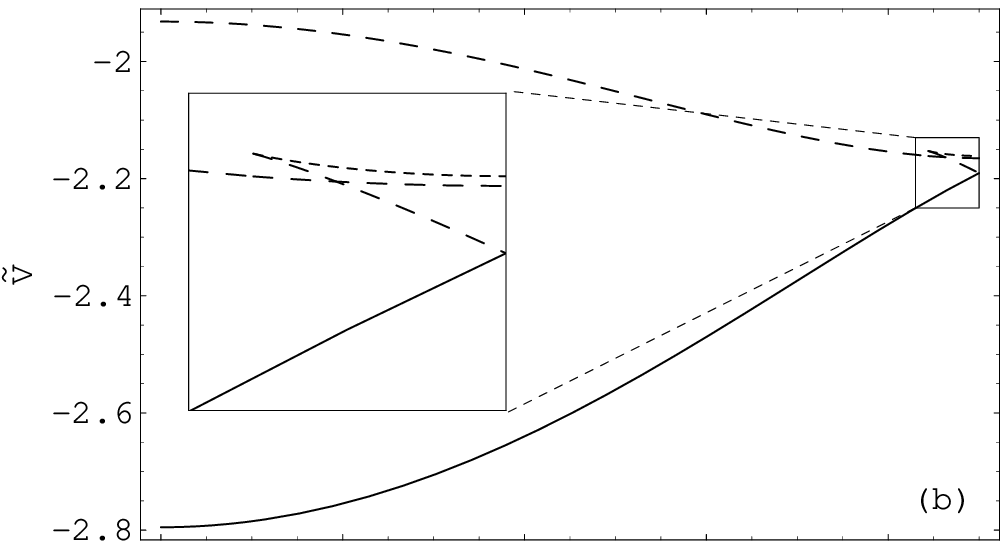}}
\centerline{\includegraphics[width=.6\textwidth]{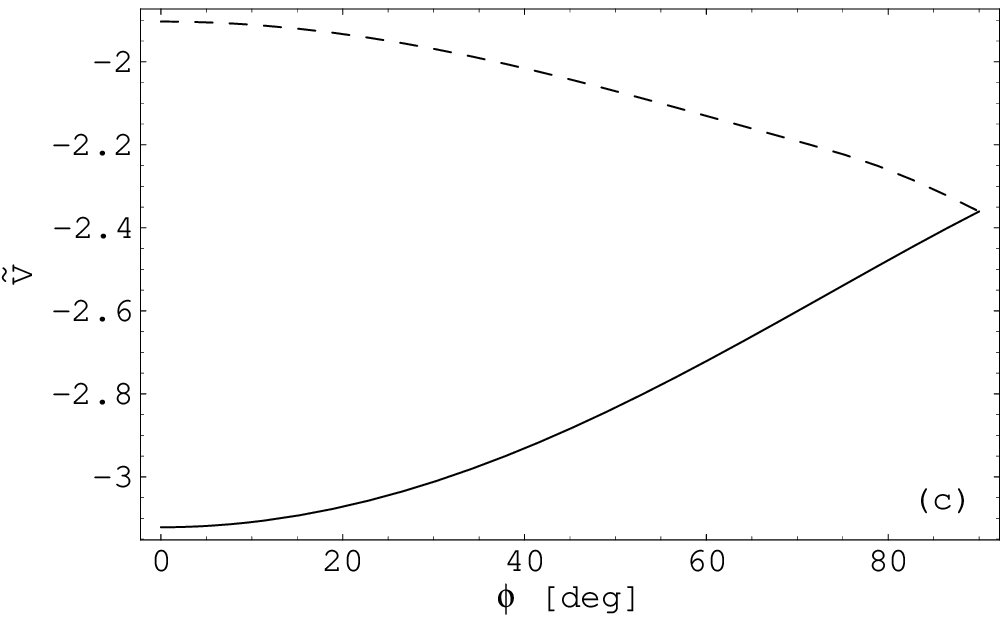}}
\caption{\label{fig:BarrAngle}
  The heights of the potential saddles as a function of the field angle
  $\phi$ for field strengths (a) $\tilde F=0.5$, (b) $\tilde F=0.75$, (c)
  $\tilde F=0.9$. Solid lines: outer saddle, long-dashed lines: inner
  saddles of index one, short-dashed lines: saddles of index two. In (b),
  the inset shows a blow-up view of the bifurcation region.}
\end{figure}

For arbitrary field angles $\phi$, the heights of the saddle points
interpolate between the two extremal cases we have just discussed. Their
angle dependence is displayed in figure~\ref{fig:BarrAngle} for three
different values of the scaled field strength. The most significant feature
the heights exhibit is that for all field strengths the height of the outer
saddle increases as the field angle increases from zero to $90^{\circ}$,
whereas the height of the inner saddle decreases. Since we have seen in
section~\ref{ssec:PPerp} that it is the outer saddle that is most relevant
for ionization, it follows that the molecule is easiest to ionize in the
parallel configuration and that ionization gets harder the larger an angle
the external field makes with the molecular axis.

In accordance with what can be seen in figure~\ref{fig:Height90}, the
behaviour of the saddle heights in the vicinity of $\phi=90^{\circ}$ is
completely different for high and for low field strengths. At low field
strength $\tilde F<\tilde F_1$, as in figure~\ref{fig:BarrAngle}(a), an
inner and an outer saddle exist which are non-degenerate. If the field
strength is increased beyond $\phi=90^{\circ}$, the saddle heights
symmetrically retrace their paths to reach their values at $\phi=0^{\circ}$
again at $\phi=180^{\circ}$. For this reason, they must have zero slope at
$\phi=90^{\circ}$.

For high field strengths $\tilde F>\tilde F_2$, as in
figure~\ref{fig:BarrAngle}(c), the two saddles are degenerate at
$\phi=90^{\circ}$. As the angles increase beyond $90^{\circ}$, the saddles
change their roles, the saddle that was ``outer'' for $\phi<90^{\circ}$
being ``inner'' for $\phi>90^{\circ}$ and vice versa. It is therefore
possible for the heights of these saddles to intersect at $\phi=90^{\circ}$
with non-zero slope. In the intermediate field strength regime $\tilde
F_1<\tilde F<\tilde F_2$, which is illustrated in
figure~\ref{fig:BarrAngle}(b),
both types of behaviour can be observed at the same time.

\subsection{Barrier heights as a function of the internuclear distance}
\label{ssec:BarrDist}

\begin{figure}
\centerline{\includegraphics[width=.6\textwidth]{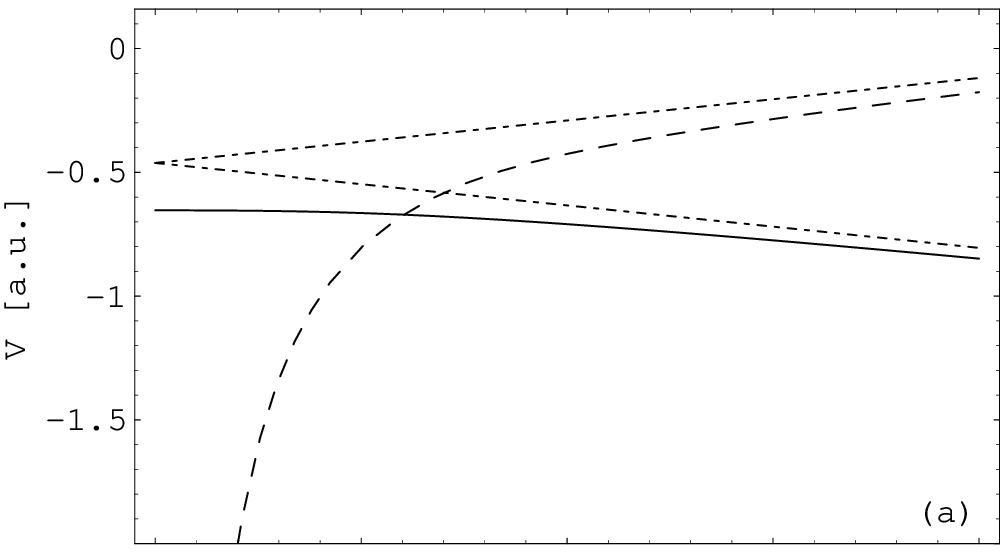}}
\centerline{\includegraphics[width=.6\textwidth]{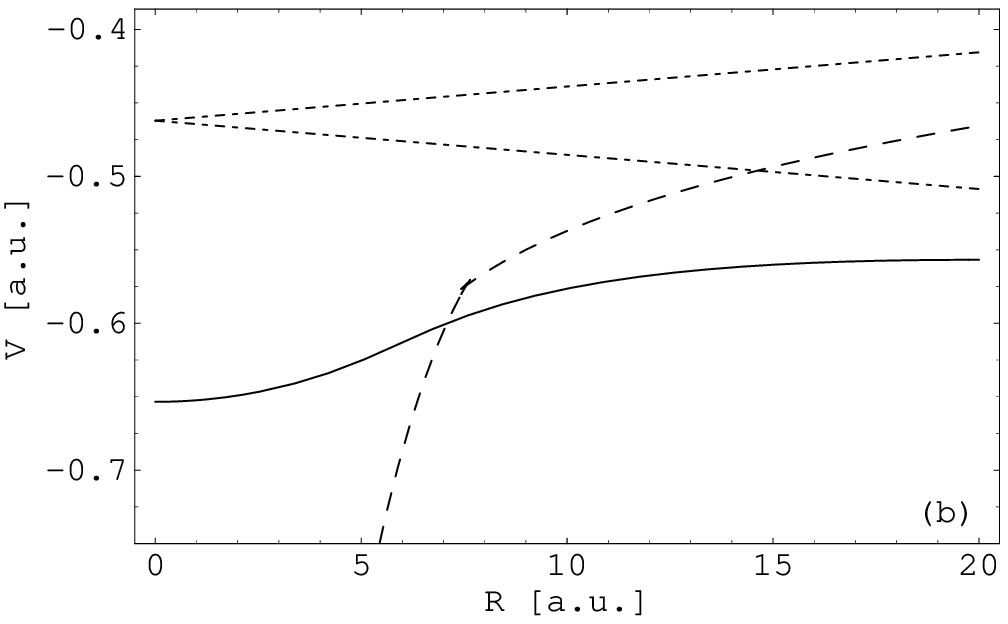}}
\caption{\label{fig:BarrDist}
  Unscaled barrier heights as a function of the internuclear separation $R$
  for fixed external electric field strength $F=0.053\,38\,{\rm a.u.}$,
  nuclear charge $Z=1$ and
  field angles (a) $\phi=50^{\circ}$, (b)
  $\phi=85^{\circ}$. Solid lines: outer saddle, dashed lines: inner
  saddles. Dash-dotted lines mark the asymptotic barrier
  heights~(\ref{SepHeight}) for
  large $R$.}
\end{figure}

So far, we have described the potential saddles as a function of the
direction and the scaled strength of the external electric field. While
this procedure facilitates the structural understanding of the saddles and
their bifurcations, it is somewhat remote from experimental conditions
where an external field of a certain non-scaled strength is given and the
value of the scaled field strength depends on the separation of the
nuclei. It can thus change considerably in time if the molecule dissociates
under the influence of the field. To approach this situation, we now turn
to a discussion of the saddle heights as a function of the internuclear
distance $R$ for fixed external field strength $F$. Results are shown in
figure~\ref{fig:BarrDist} for $F=0.053\,38\,{\rm a.u.}$, which is the peak
field strength of a linearly polarized laser field with an intensity of
$10^{14}\, {\rm W\,cm^{-2}}$. The nuclear charge was chosen to be $Z=1$, as
is appropriate for a hydrogen molecular ion.

The most conspicuous feature of figure~\ref{fig:BarrDist} is that the
height of the inner saddle diverges to $-\infty$ for $R\to 0$ whereas the
height of the outer saddle remains finite. This behaviour can be understood
by noting that the limit $R\to 0$ corresponds to the united-atom limit
where the two nuclei coincide to form a single  nucleus of double
charge. If a hydrogen-like atom of nuclear charge $z$ is exposed to an
external electric field of strength $F$, there is a single potential
``Stark''-saddle whose height is \cite{Friedrich98}
\begin{equation}
  \label{StarkSad}
  V_{\rm Stark}=-2\sqrt{zF}\;.
\end{equation}
For the field strength used in figure~\ref{fig:BarrDist}, the Stark saddle
height is $V_{\rm Stark}=0.6535\,{\rm a.u.}$, which is the value approached
by the outer saddle height. It is independent of the field angle $\phi$
because in the united-atom limit the notion of an intermolecular axis
becomes meaningless. The inner saddle, by contrast, is always located
within a few $R$ of either nucleus. Thus, in the limit $R\to 0$ it
approaches both nuclei, so that the saddle height must diverge.

In the opposite limit of large internuclear distance, the nuclei are so far
apart that their Coulomb wells barely overlap. Therefore, if an external
field is present, to each nucleus there is attached a Stark saddle point
whose height relative to the value of the external potential at the
location of the nucleus is given by~(\ref{StarkSad}). The nuclei themselves
are displaced to opposite sides from the zero of the external potential,
which we have chosen in~(\ref{Potential}) to be at the midpoint between the
nuclei. The total height of the two saddles in the separated-atom limit is
thus
\begin{equation}
\label{SepHeight}
  V_{\rm sep}=-2\sqrt{F}\pm\frac R2 F\cos\phi \;.
\end{equation}
These asymptotic saddle heights are indicated by the dash-dotted lines in
figure~\ref{fig:BarrDist}.  From 
figure~\ref{fig:TccField}(b) it becomes clear that the inner saddle is
shifted into the direction of higher external potential, the outer saddle
into the direction of lower potential. Thus, the inner saddle height must
approach the higher of the two asymptotic values~(\ref{SepHeight}) while
the outer saddle height approaches the lower value.

Due to the asymptotic behaviour~(\ref{SepHeight}), the height of the inner
saddle increases with $R$ for large internuclear distances, the height
of the lower saddle decreases. However, the distance-dependence of the
asymptotic heights~(\ref{SepHeight}) becomes slight when the angle
approaches $90^{\circ}$, so that the asymptotic behaviour sets in only for very
large distances. Thus, for large angles we find that the outer barrier
height is actually increasing with the internuclear separation over the
physically relevant range of distances.

\begin{figure}
\centerline{\includegraphics[width=.6\textwidth]{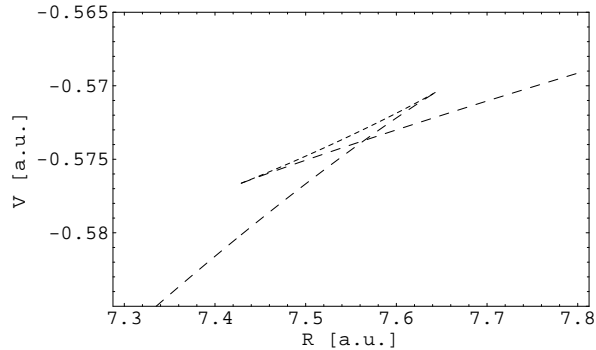}}
\caption{\label{fig:DistZoom}
  Enlarged detail of figure~\ref{fig:BarrDist}(c), showing the bifurcations
  of the inner saddles. Long-dashed lines: saddles of index one,
  short-dashed line: saddle of index two.}
\end{figure}

At first sight, the bifurcations of the inner saddles seem to have no
impact whatsoever on the behaviour of the barrier heights in
figure~\ref{fig:BarrDist}. Upon a closer look, however, one discovers a
little ``knee'' in the inner saddle height for $\phi=85^{\circ}$ at
$R\approx 7.5\,{\rm a.u}$. At this point the curve seems to change its
slope discontinuously. If we enlarge this part of the curve, as shown in
figure~\ref{fig:DistZoom}, we find that this is indeed the region where the
bifurcations take place. In a small range of internuclear distances, a new
saddle of index one is created and the previously-existing saddle is
destroyed. This shift from one saddle to the other gives rise to the
apparently discontinuous bend observed in figure~\ref{fig:BarrDist}(b).

\section{Tunnelling ionization}
\label{sec:Tunnel}

Even if the electron energy is below the barrier height discussed in
section~\ref{sec:Thresh}, ionization can still take place due to quantum
mechanical tunnelling. In atoms, this process is commonly described in the
framework of the ADK model \cite{Ammosov86}, which uses the ionization rate
of a hydrogen atom or a hydrogen-like ion in a static external field. In
that system, ionization takes place predominantly along the axis of the
external electric field, so that the description of the ionization process
reduces to an essentially one-dimensional tunnelling calculation which
matches the wave function within the well to the ionizing parts of the wave
function beyond the barrier. To generalize this approach to the ionization
of molecules, a double modification is necessary: On the one hand, a
molecular wave function is anisotropic, so that the ionization rate will
depend on the direction in which ionization is taking place, i.e.~on the
direction of the external field. On the other hand, in the complicated
potential~(\ref{Potential}), unlike for the hydrogen atom in an electric
field \cite{Friedrich98,LandauI}, the different degrees of freedom do not
decouple, so that the tunnelling process cannot be regarded as essentially
one-dimensional. This coupling of the different degrees of freedom must
also be taken into account.

Recently, Tong \etal \cite{Tong02} proposed a generalization of the ADK
model of tunnelling ionization to the ionization of diatomic
molecules. They take into account the anisotropy of the bound-state
molecular wave function, but they retain an essentially one-dimensional
picture of the tunnelling process and fail to discuss the impact of
multiple coupled degrees of freedom. We will here analyze the tunnelling
paths for the hydrogen molecular ion in a non-axial electric field and
thereby elucidate the reason why the quasi one-dimensional approach of Tong
\etal could succeed even though the effects of the additional degrees of
freedom were neglected.

Different approaches to the theory of barrier-suppression ionization are
surveyed by Kra{\u\i}nov \cite{Krainov95}.  We will here adopt a
semiclassical description of the ionization process. Within that framework,
the tunnelling transmission probability through a barrier~\cite{LandauIII}
\begin{equation}
  \label{TunnProb}
  P=\frac{1}{1+\rme^{K/\hbar}}
\end{equation}
is determined by the action integral
\begin{equation}
  \label{TunnAct}
  K=\int |p|\,dq
\end{equation}
along the tunnelling path, with
\begin{equation}
  \label{TunnMomentum}
  p({\bi x})=\sqrt{2(E-V({\bi x}))}
\end{equation}
the (imaginary) momentum the electron has at the position $\bi x$ below the
barrier, $V({\bi x})>E$. Kra{\u\i}nov \cite{Krainov95} gives a formula for
the tunnelling probability in the case of a hydrogen atom in an electric
field that is obtained from~(\ref{TunnProb}) if the barrier is assumed
parabolic. In its more general form, equation~(\ref{TunnProb}) is
applicable to parabolic as well as non-parabolic barriers and in one as
well as several degrees of freedom.  In one dimension, the
action~(\ref{TunnAct}) is computed along the unique path leading from one
side of the barrier to the other. In several dimensions, however, an
infinity of possible tunnelling paths connect the region of bound motion at
a given energy to the region of free motion, and it is not obvious along
what path the tunnelling action~(\ref{TunnAct}) should be taken. A similar
difficulty arises if the strictly quantum mechanical method of the phase
functions \cite{Krainov95} is used in multiple degrees of freedom. Thus, we
expect the results described below to be valuable beyond the limits of the
semiclassical theory.

If reactant and product regions of configuration space are separated by a
saddle point in a multidimensional potential, the path of steepest descent
from the saddle is regarded as the ``reaction path'' in chemistry (see,
e.g., \cite{Truhlar96}). However, it is also well known that the tunnelling
integral taken along that path can severely underestimate the extent of
tunnelling.  To obtain the transmission probability correctly, the path
must be chosen such as to maximize~(\ref{TunnProb}), i.e.~to minimize the
tunnelling action~(\ref{TunnAct}). In cases where the reaction path is
strongly curved, the tunnelling path lies on its concave side (``corner
cutting'') \cite{Child91},
which shortens the path and helps minimize the tunnelling action.

Through~(\ref{TunnMomentum}) the optimum tunnelling path depends
crucially on the details of the potential, or equivalently the dynamics,
in the vicinity of the saddle point. To analyze it, it is helpful to allow
coordinates, momenta and the time to assume complex rather than real
values. This artifice allows us extend the classical dynamics to the region
below the barrier and makes the full machinery of classical dynamics
available to an analysis of the tunnelling process. If we do so, by virtue
of Maupertuis' principle \cite{LandauI} the requirement to minimize the
tunnelling action~(\ref{TunnAct}) translates into the condition that the
tunnelling path be the configuration space projection of a classical
trajectory. In addition, the tunnelling path must start and end on the
energy contour $V({\bi x})=E$, where the real dynamics can take over. At
these points, it must have zero momentum. Using the time-reversal
invariance of the dynamics described by the Hamiltonian~(\ref{Ham}), we can
conclude that the tunnelling orbit must be periodic. This leads us to the
prescription to identify the optimum tunnelling path as the configuration
space projection of an imaginary-time complex periodic orbit under the
potential barrier. Equivalently, it can be found as the projection of a
real periodic orbit in the inverted potential \cite{Miller75}.

\begin{figure}
\centerline{\includegraphics[width=.48\textwidth]{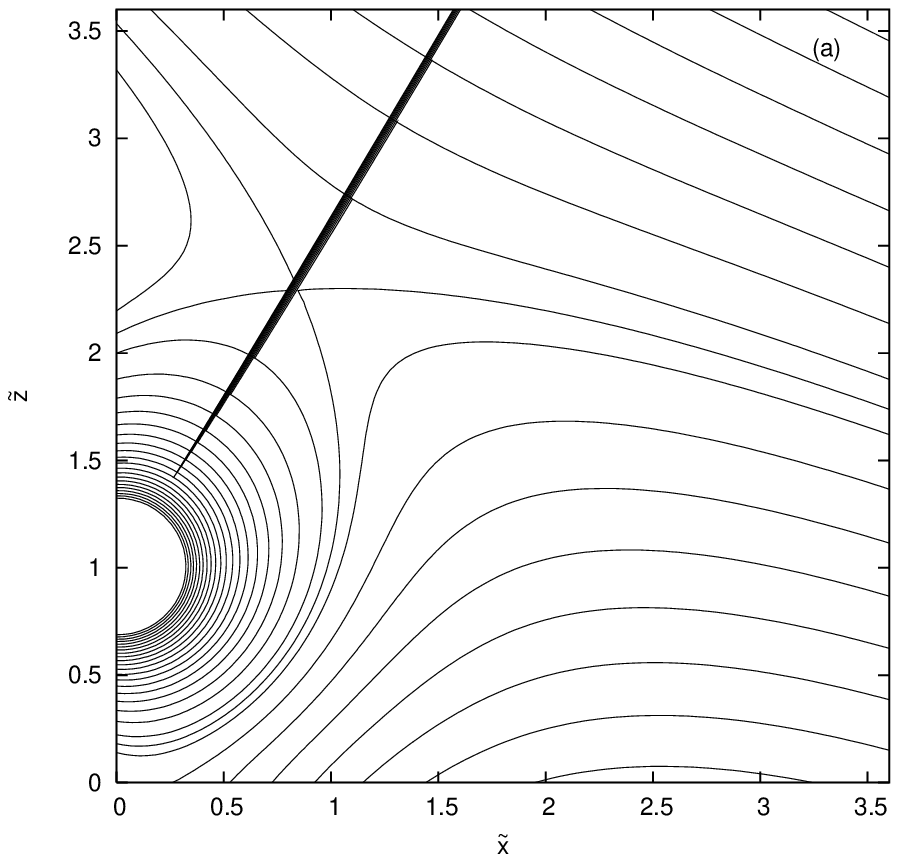}\hfill
            \includegraphics[width=.48\textwidth]{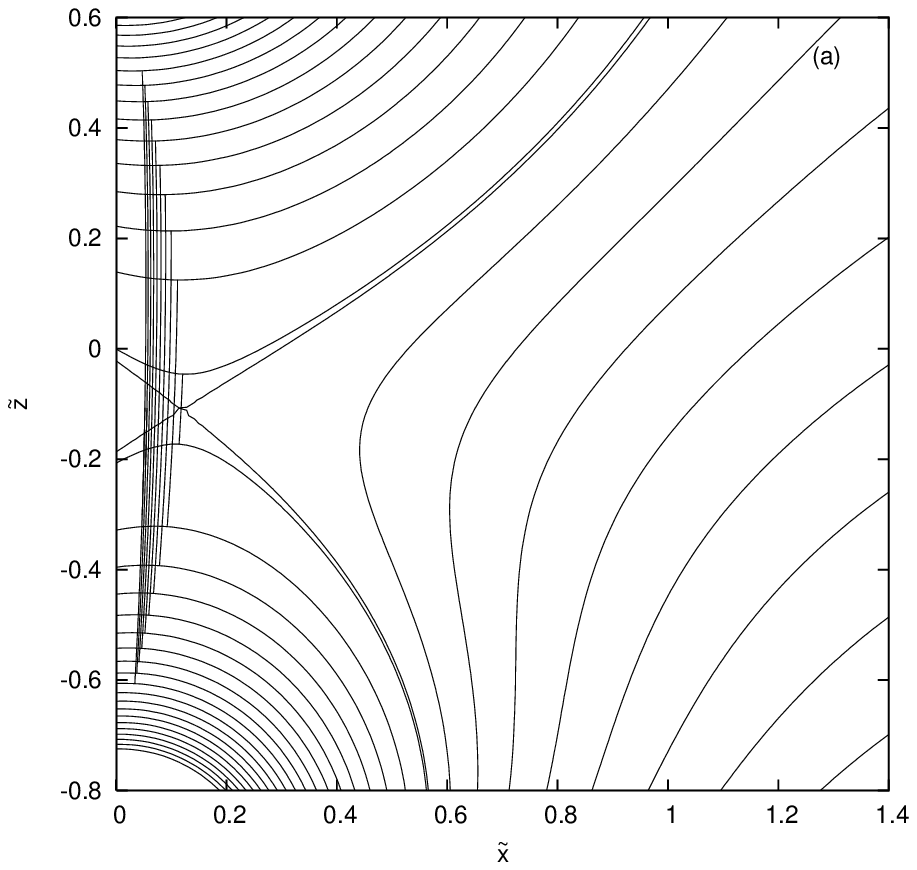}}
\centerline{\includegraphics[width=.48\textwidth]{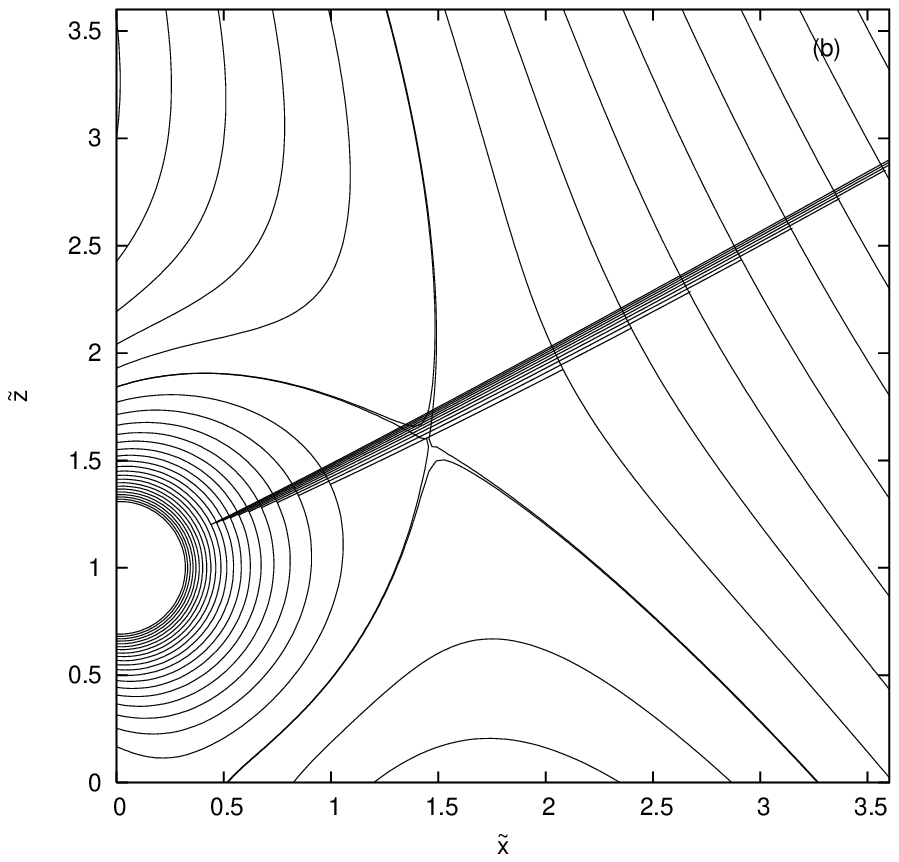}\hfill
            \includegraphics[width=.48\textwidth]{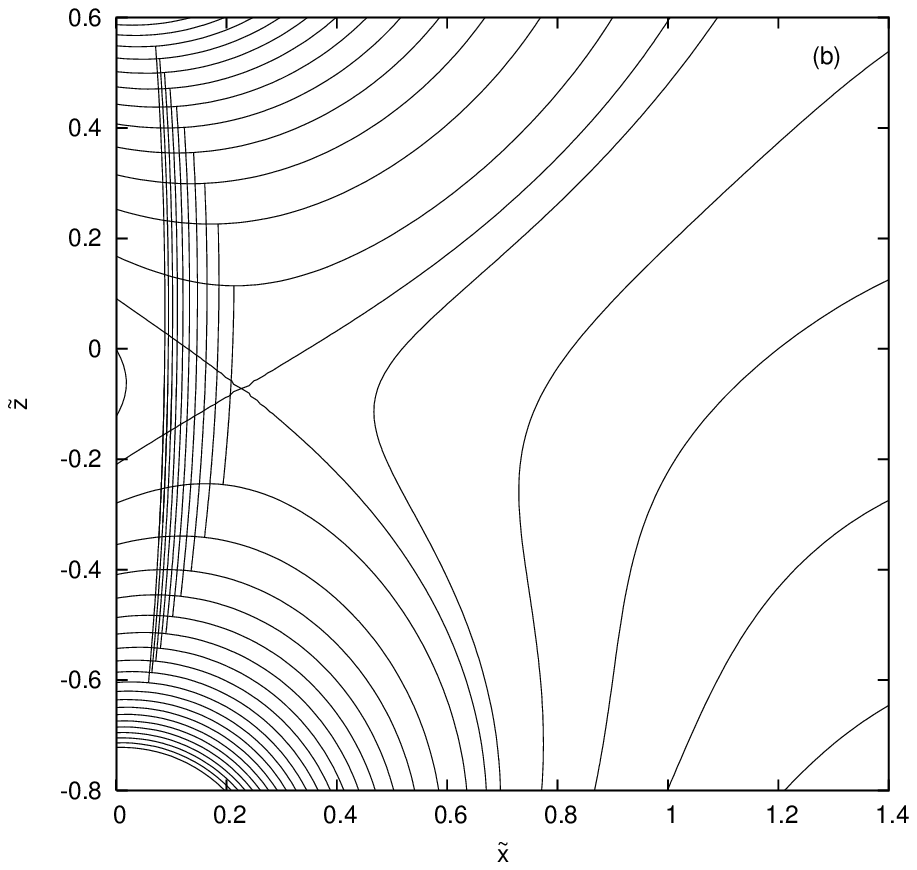}}
\centerline{\includegraphics[width=.48\textwidth]{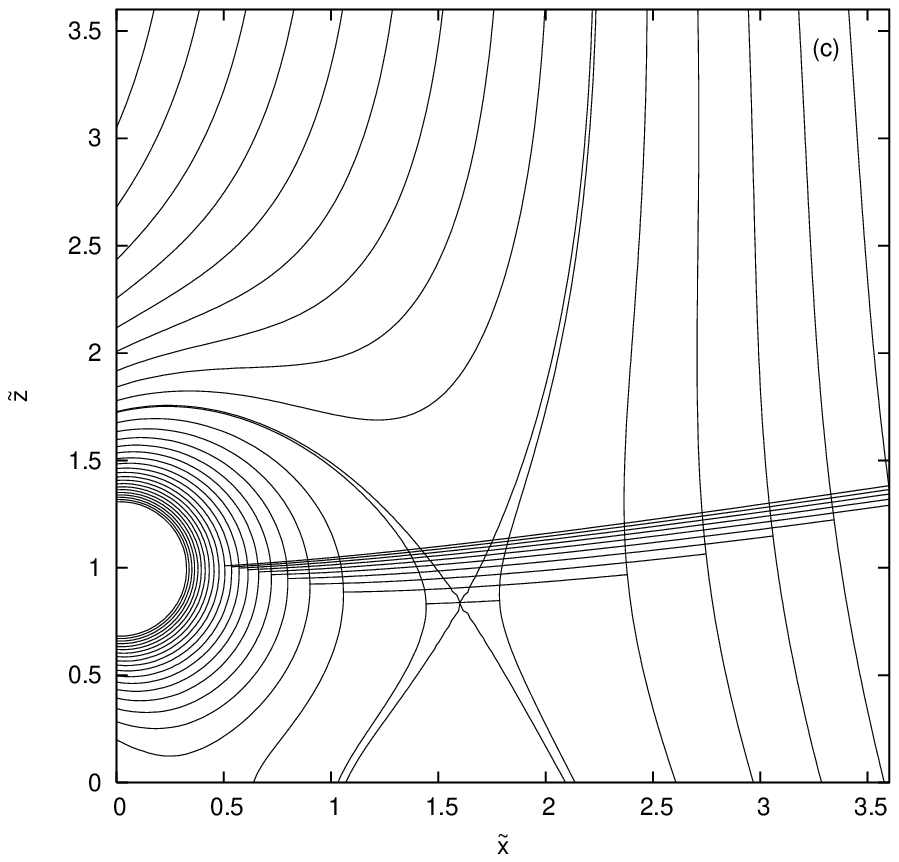}\hfill
            \includegraphics[width=.48\textwidth]{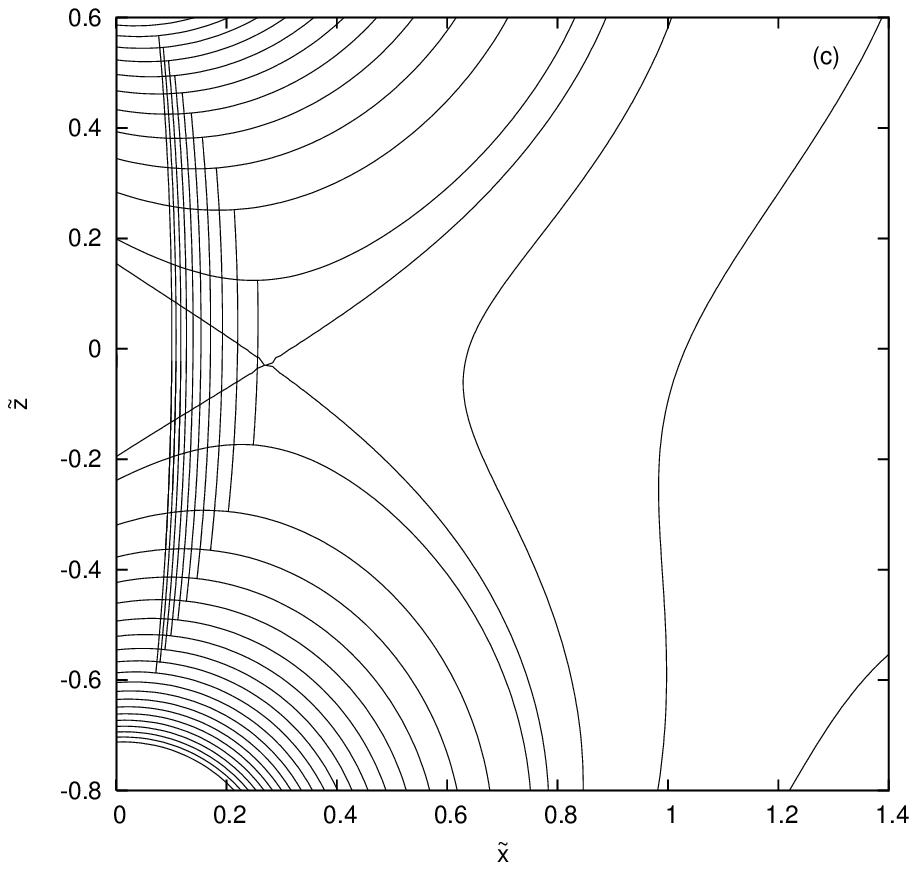}}
\caption{\label{fig:Path}
  Optimum tunnelling paths for tunnelling under the outer (left column) and
  inner (right column) saddles at different energies for
  a scaled external electric field strength
  of $\tilde F=0.5$ and field angles (a) $\phi=30^{\circ}$, (b)
  $\phi=60^{\circ}$, and (c) $\phi=80^{\circ}$.}
\end{figure}

Using the above prescription, we can compute the tunnelling paths for
arbitrary external field strengths, field directions and tunnelling
energies. Results for a scaled external field strength $\tilde F=0.5$ are
shown in figure~\ref{fig:Path}. These tunnelling paths show the expected
shift to the concave side of the reaction path, but the amount of corner
cutting is small because, especially for small angles and for the outer
saddle, the reaction path is close to a straight line. This is a clear
indication of quasi one-dimensional tunnelling dynamics.

\begin{figure}
\centerline{\includegraphics[width=.6\textwidth]{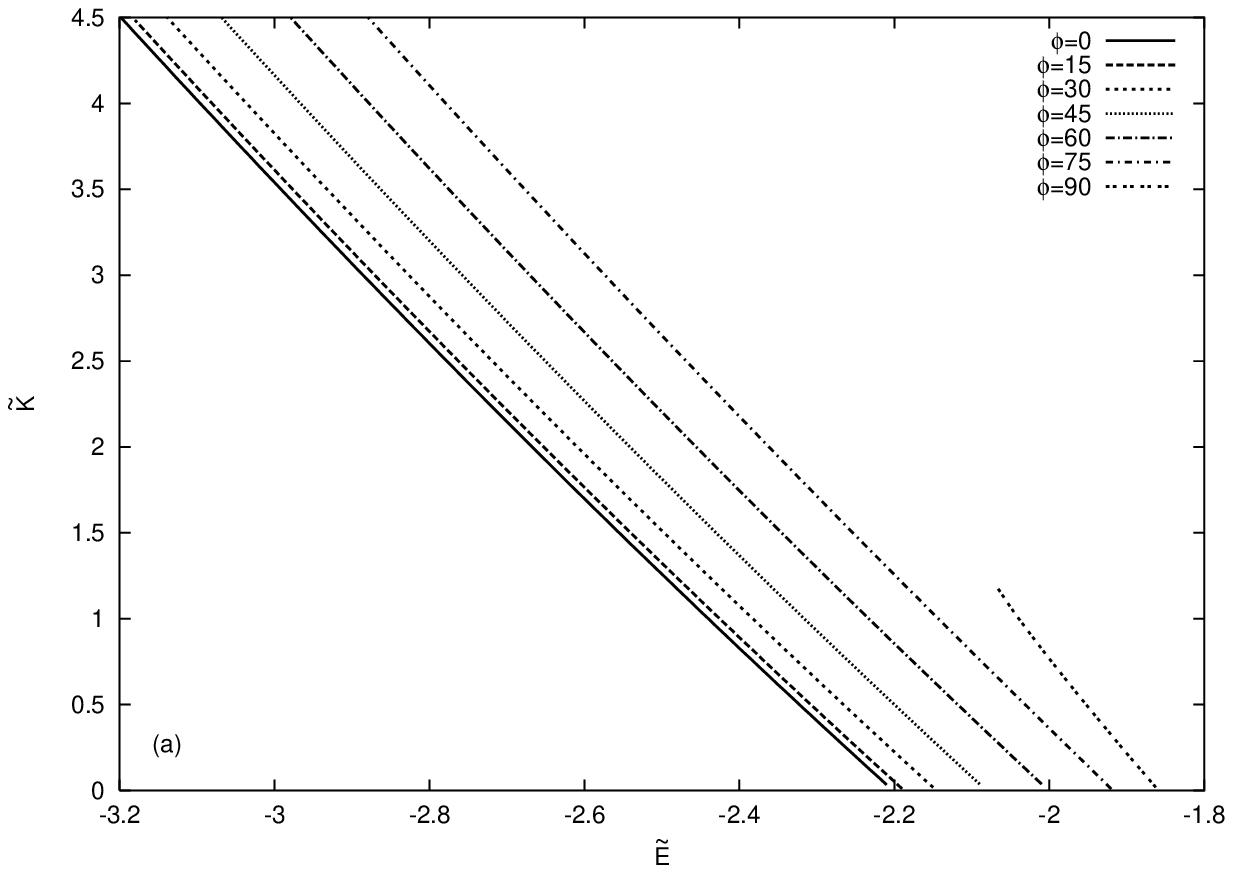}}
\centerline{\includegraphics[width=.6\textwidth]{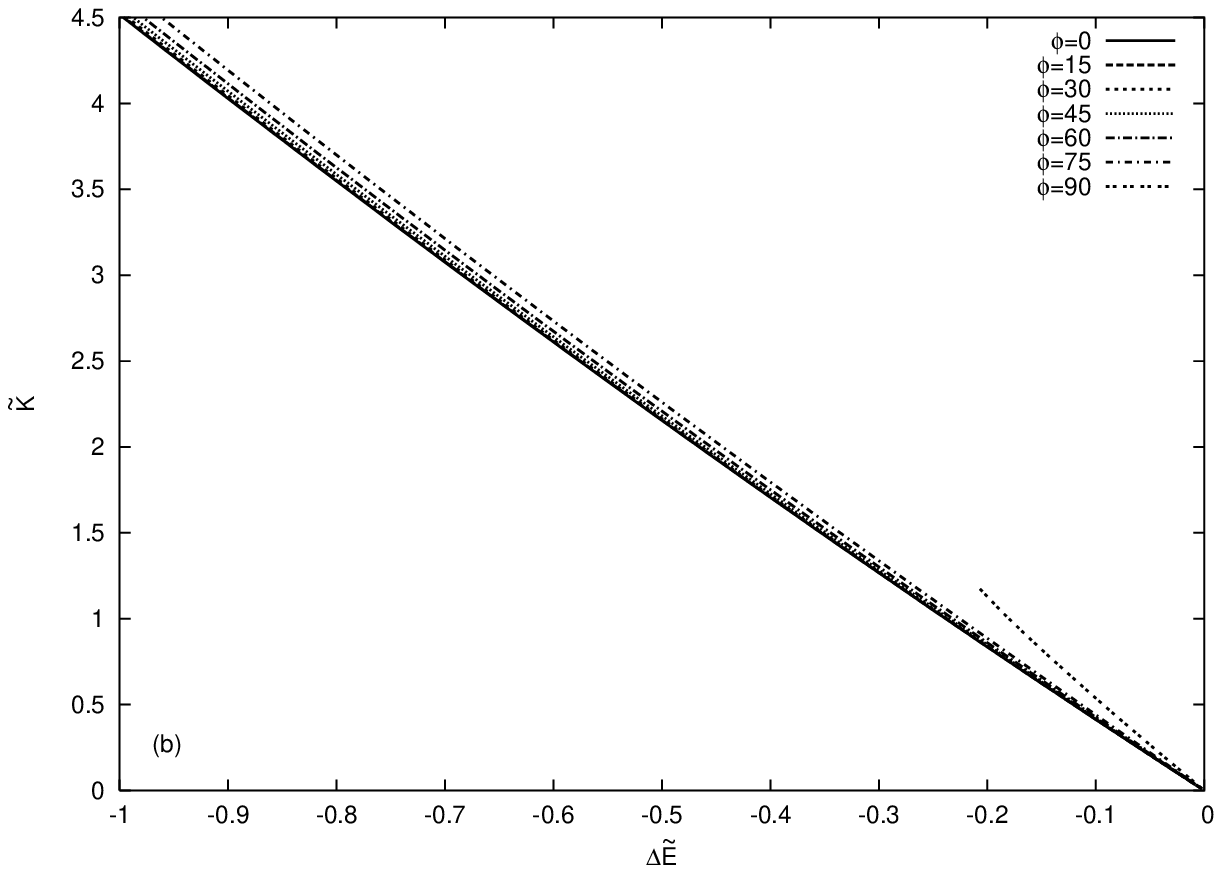}}
\caption{\label{fig:Action}
   Tunnelling actions for scaled electric field
  strength $\tilde F=0.5$ and different field angles, plotted as a function
  of absolute scaled energy (a) and scaled energy difference below the
  height of the saddle point (b).}
\end{figure}

Even more striking is the behaviour of the tunnelling actions shown in
figure~\ref{fig:Action}. The curves in figure~\ref{fig:Action}(a) present
the energy-dependence of the tunnelling action below the outer saddle for
different angles. They differ mainly in the height of the saddle point in
which they originate. This is especially clear when all energies are
referred to the respective saddle point energies, as in
figure~\ref{fig:Action}(b). In this case, all action curves nearly
coincide. An exception is formed by the curve for $\phi=90^{\circ}$. This
curve is lying on the $x$-axis, as are both saddle points. If the energy is
below the height of the inner saddle, the periodic tunnelling orbit will
cross that saddle and the tunnelling path described above will cease to
exist. The presence of the second saddle distorts the action curve away
from the universal behaviour of the other curves and finally causes it to
end.

The nearly universal tunnelling dynamics is due to the fact that the outer
saddle, especially for small field angles, is much closer to one of the
nuclei than to the other. Thus, the neighbourhood of the saddle is very
similar to that of the Stark saddle encountered in a single atom in an
electric field. The tunnelling dynamics is therefore similar the atomic
tunnelling dynamics, and the ionization rate is closely approximated by the
ADK rate once the distortion of the molecular wave function through the
presence of the second nucleus is taken into account. The tunnelling rate
will, however, depend on the field angle through the height of the saddle.
Very similar results are found for the higher scaled field strength
$\tilde F=0.75$.

\begin{figure}
\centerline{\includegraphics[width=.6\textwidth]{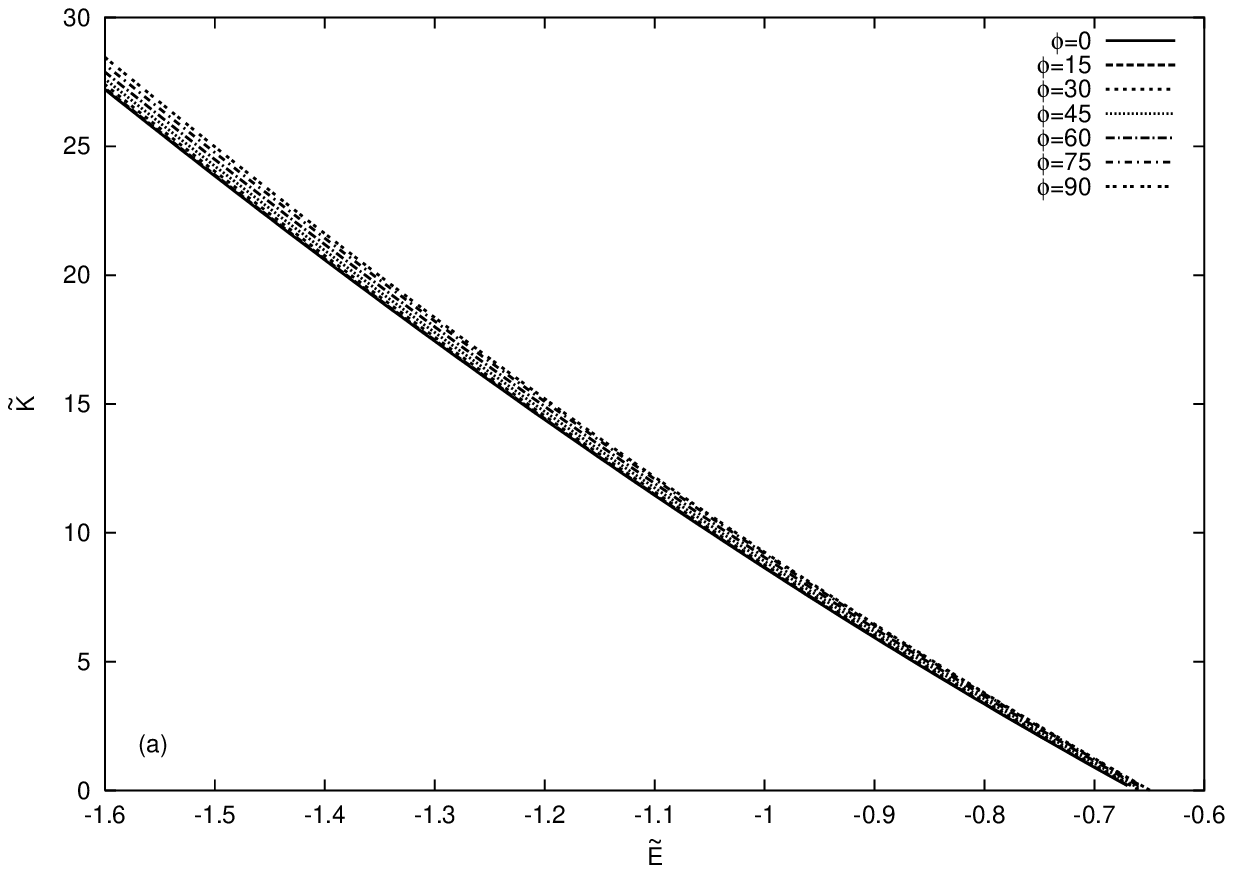}}
\caption{\label{fig:Action2}
  Tunnelling actions for scaled electric field
  strength $\tilde F=0.05338$ and different field angles, plotted as a function
  of scaled energy.}
\end{figure}

The universality of the tunnelling dynamics is even more pronounced for the
lower scaled field strength $\tilde F=0.053\,38$, which is the peak field
strength in a linearly polarized laser beam with an intensity of
$10^{14}{\rm W\,cm}^{-2}$. In this case, the inner
saddle point is very close to the origin $\tilde x=\tilde z=0$, so that the
inner tunnelling paths lie virtually on the $\tilde z$-axis. The outer
saddle is located at a scaled distance from the origin between 6.35 at
$\phi=0$ and 6.00 at $\phi=90^{\circ}$. For these large distances, we
approach the united-atom limit, which explains why the angle dependence of
the distance is slight. As a further consequence, both the saddle shapes
and saddle heights are well approximated by the Stark saddle of the united
atom, as is confirmed by the tunnelling actions shown in
figure~\ref{fig:Action2}. We find not only that the outer saddle height is
virtually independent of the field angle, but also that the tunnelling
action is a nearly universal function of the tunnelling energy for all
angles including $\phi=90^{\circ}$. At this field strength, the inner
saddle is sufficiently far away from the outer saddle so that it does not
cause any distortion or singularity in the energy range shown.

The experimental data used by Tong \etal~\cite{Tong02} was taken at laser
intensities around $I=10^{14}\,{\rm W\,cm^{-2}}$, which corresponds to a
peak field strength of $F=0.053\,38\,{\rm a.u.}$ and, for an internuclear
separation of $2\,{\rm a.u.}$, to the scaled field strength of $\tilde
F\approx 0.053\,38$ we have studied above. It is well within the
range of field strengths where the angle-independent behaviour of the
united-atom limit holds. For a neutral molecule, in particular a
non-hydrogenic molecule as in~\cite{Tong02}, the effective potential
experienced by the ionizing electron is more complicated than the
potential~(\ref{Potential}) due to the presence of the inner
electrons. However, these deviations from Coulombic behaviour arise mainly
in the vicinity of the nuclei.  For the (on an atomic scale) weak fields
relevant to~\cite{Tong02}, the saddle is far away from the nuclei and we
can expect the simple potential~(\ref{Potential}) to approximate the true
tunnelling dynamics well.

We can therefore conclude that for small field strengths the effects of
reaction path curvature that were not considered by Tong \etal have little
impact on the tunnelling ionization of a molecule because the
electron is essentially subject to the Coulomb field of the united
nuclei. They will become even smaller when averaged over the
external field direction. This observation explains why the quasi-atomic
tunnelling theory of~\cite{Tong02} could succeed without taking these
dynamical effects into account. At the same time, the angle-independence of
the tunnelling dynamics explains why previous theories of ionization that
only deal with the special case of an axial field
\cite{Codling89,Zuo95,Posthumus95,Posthumus96a,Plummer96,Smirnov97,Smirnov98}
yield results in resonable agreement with experiment although in an
experiment one has no control over the relative orientations of the
molecule and the field.

In the opposite limit of strong external field, we regain a similar
universal tunnelling dynamics because the saddle points approach individual
nuclei and the influence of the remote nucleus becomes small. As we have
found only small deviations from universal quasi-atomic dynamics even for
intermediate field strength, we can conclude that the effects of reaction
path curvature are small for all field strengths.

\section{Conclusion}

As a first step toward an understanding of the electronic dynamics of the
hydrogen molecular ion in an electric field of arbitrary orientation, we
have presented a detailed discussion of the saddle points of the relevant
potential. We have found that the saddle points bifurcate as the external
electric field strength and direction are varied, which gives us a first
glimpse of the richness the full dynamics of the system must possess.  We
used our knowledge of the saddle points to describe both the onset of
over-the-barrier ionization and the dynamics of tunnelling ionization below
the saddle points, which shed new light on the success of a recent theory
of molecular tunnelling ionization~\cite{Tong02} as well as previous
theories that treat the special case of axial fields
\cite{Codling89,Zuo95,Posthumus95,Posthumus96a,Plummer96}.

The presentation of the dynamics near the saddle points in the hydrogen
molecular ion opens several directions of further investigation: Firstly,
it calls for a more thorough investigation of the dynamics within the
wells. One can expect to find a wealth of intricate classical structure
that could illuminate the finer details of the molecular ionization
process, both above and below the classical threshold. Secondly, the
connection to quantum mechanics must be established to determine the
energies and widths of the bound states within the well in a way analogous
to the one-dimensional calculations of
\cite{Connor81,Connor82,Connor82a}. Thirdly, with regard to Rydberg plasmas
it is essential to include the Coulomb fields of the surrounding ions and
to allow for the presence of a magnetic field.  The dynamics of the
hydrogen molecular ion in an electric field of arbitrary direction offers a
rich and fascinating field of inquiry whose investigation we have here only
just begun.

\section*{Acknowledgments}
TB is grateful to the Alexander von Humboldt-Foundation for a
Feodor Lynen fellowship. TU acknowledges financial support from the
National Science Foundation.

\section*{References}

\end{document}